\documentclass[aps,prb]{revtex4-1}

\usepackage{graphicx}
\usepackage{siunitx}
\usepackage{amsmath}
\usepackage{amsfonts}
\usepackage{amssymb}
\usepackage{epsfig}
\usepackage{graphicx}

\newcommand{\beq}{\begin{equation}}
\newcommand{\eeq}{\end{equation}}
\newcommand{\beqa}{\begin{eqnarray}}
\newcommand{\eeqa}{\end{eqnarray}}

\begin{document}

\title{Power Spectrum and Diffusion of the Amari Neural~Field}

\author{Luca Salasnich}  

\address{Dipartimento di Fisica e Astronomia 
``Galileo Galilei'' and CNISM, 
Universit\`a di Padova, Via Marzolo 8, 35131 Padova, Italy; luca.salasnich@unipd.it \\   
Istituto Nazionale di Ottica (INO) del Consiglio Nazionale 
delle Ricerche (CNR), Via Nello Carrara 1, \mbox{I-50019 Sesto Fiorentino, Italy}}

\begin{abstract}
We study the power spectrum of a space-time dependent 
neural field which describes the average membrane potential of neurons 
in a single layer. This neural field is modelled by a dissipative 
integro-differential equation, the so-called Amari equation. 
By considering a small perturbation with respect to a stationary and 
uniform configuration of the neural field we derive a linearized 
equation which is solved for a generic external 
stimulus by using the Fourier transform into wavevector-freqency domain,  
finding an analytical formula for the power spectrum of the 
neural field. In addition, after proving that for large 
wavelengths the linearized Amari equation is equivalent 
to a diffusion equation which admits space-time dependent 
analytical solutions, we take into account the nonlinearity 
of the Amari equation. We find that for large wavelengths 
a weak nonlinearity in the Amari equation 
gives rise to a reaction-diffusion equation 
which can be formally derived from a neural action functional 
by introducing a dual neural field. 
For some initial conditions, we discuss 
analytical solutions of this reaction-diffusion equation.
\end{abstract}

\maketitle

\section{Introduction}

Neural field theory is the set of models of brain organization 
and function in which the interaction of billions of neurons 
is treated as a continuum \cite{review-bressloff,book-nft}. 
It was Wilson and Cowan 
\cite{wilson}, Nunez \cite{nunez}, and Amari \cite{amari} in the 1970s 
who provided the formulations for neural field models that are in common 
use today \cite{book-nft}. 

In this paper we analyze one of the most used formulations 
of neural field activity: The deterministic Amari's equation \cite{amari}, 
which describes the local activity of a population of neurons 
in a single-layer. We linearize the Amari equation and Fourier-transform it 
from the space-time domain to the wavevector-freqency domain. In this way  
we obtain an elegant analytical solution of the equation and, 
in particular, we determine the power spectrum of the neural field in the case 
of an instantaneous 
and localized external stimulus. 
In the regime of large frequency we show that 
the power spectrum scales as $1/\omega^2$, which is indeeed 
a direct consequence of the exponential decay in the time domain. 
The same $1/\omega^2$ law has been obtained \cite{buice1,buice2} 
investigating the stochastic version of the Wilson-Cowan neural 
field theory \cite{buice2}. This result is also consistent with the scaling 
laws found in measurements of electroencephalography (EEG) \cite{eeg1}. 
Clearly, EEG spectra of intact functional brains are quite complex, 
showing very prominent resonances (alpha, beta, etc) beyond the 
$1/\omega^2$ shoulder~\cite{eeg2}. Modelling these resonances is 
one of the central issue of theoretical neuroscience \cite{robinson}. 

In addition, in this paper we find that for small wavenumbers 
the Fourier antitransform of the linearized Amari equation gives 
a diffusion equation, which is thus reliable for large wavelengths. 
Finally, we investigate some consequences of nonlinearity 
in the Amari equation. For large wavelengths and weak nonlinearity 
we deduce a reaction-diffusion equation and discuss some of its 
spatially uniform solutions. 
We show that this diffusion equation can be obtained 
by extremizing a neural action functional by introducing a dual 
neural field. This action functional is very 
similar to a neural action recently obtained \cite{buice2} 
within a stochastic extension of the Wilson-Cowan model. 

\section{Amari Equation}

The Amari equation \cite{amari} is given by 
\beq 
\tau \, {\partial \over \partial t} u({\bf r},t)= - u({\bf r},t) + 
\int d^d{\bf r}' \, w({\bf r}',{\bf r}) f[u({\bf r}',t)] + s({\bf r},t) \; , 
\eeq
where $u({\bf r},t)$ is the space-time dependent neural field, i.e.,  
the average membrane potential of neurons 
at the position ${\bf r}$ and time $t$. Usually $d=3$ but 
often one works with $d=1$ or $d=2$ \cite{review-bressloff,book-nft}. 
Here $\tau$ is the (constant) decay time 
of the one-layer membrane of neurons, $w({\bf r}',{\bf r})$ is the 
synaptic connection weight from a position ${\bf r}'$ to another 
position ${\bf r}$. We assume that the connections are simmetric 
\beq 
w({\bf r}',{\bf r}) = w(|{\bf r}'-{\bf r}|) \; . 
\label{amari}
\eeq

The nonlinear function $f[u]$ is the activation function usually 
modelled as a sigmoid, i.e.,  a Fermi-Dirac distribution 
\beq 
f[u] = {1\over e^{\beta (h - u)} + 1} \;  
\label{fermi}
\eeq 
with gain $\beta >0$ and threshold $h>0$. Finally, 
$s({\bf r},t)$ is an external stimulus 
acting on neurons \cite{review-bressloff,book-nft,amari}. 

Let us suppose that the exteral stimulus is absent, i.e.,  $s({\bf r},t)=0$. 
It is clear from Equation~(\ref{amari}) that a stationary and uniform 
configuration $u_0$ of the neural field $u({\bf r},t)$ 
satisfies the nonlinear algebric~equation 
\beq 
u_0 = W \, f[u_0] \; , 
\label{constant}
\eeq
where 
\beq 
W = \int d^d{\bf r} \, w(|{\bf r}|) \; . 
\eeq

If $u_0\ll h$ from Equation (\ref{fermi}) one finds 
$f[u_0] \simeq 0$ and consequently from Equation (\ref{constant}) 
it follows $u_0 \simeq 0$. Instead, if $u_0 \gg h$ from Equation (\ref{fermi}) 
one finds $f[u_0] \simeq 1$ and consequently from 
Equation (\ref{constant}) it follows $u_0 \simeq W$ if $W>h$ or $u_0\simeq 0$ 
if $W<h$. 

\section{Linearized Amari Equation}
 
We now consider a perturbation $\eta({\bf r},t)$ with respect 
to the configuration $u_0$ of the neural field,~namely 
\beq
u({\bf r},t) = u_0 + \eta({\bf r},t) \; . 
\eeq

If the neural perturbation is sufficiently small, 
i.e.,  $|\eta({\bf r},t)|\ll u_0$, we can write 
\beq 
f[u({\bf r}',t)] \simeq f[u_0] + f'[u_0] \eta({\bf r}',t) 
\label{expand}
\eeq
and Equation (\ref{amari}) gives 
\beq 
\tau \, {\partial \over \partial t} \eta({\bf r},t)= - \eta({\bf r},t) + 
f'[u_0] \int d^d{\bf r}' \, w(|{\bf r}'-{\bf r}|) \, \eta({\bf r}',t) 
\label{linear-amari}
\eeq
taking into account Equations (\ref{constant}) and (\ref{expand}). This is 
the linearized Amari's equation around a uniform and constant 
configuration $u_0$. 

\subsection{Power Spectrum of the Linearized Amari Equation}

Equation (\ref{linear-amari}) can be transformed into an algebric equation 
by introducing the Fourier transform~\cite{fourier-book}
\beq 
{\tilde \eta}({\bf k},\omega) = {\cal F}[\eta({\bf r},t)]({\bf k},\omega) 
\eeq
where 
\beq 
{\cal F}[\eta({\bf r},t)]({\bf k},\omega) = \int d^d{\bf r} \, dt \, 
\eta({\bf r},t) \, e^{-i({\bf k}\cdot {\bf r}-\omega t)} \; , 
\eeq
with ${\bf k}$ the wavevector and $\omega$ the frequency and, by definition, 
\beq 
\eta({\bf r},t) = {1\over (2\pi)^{d+1}}
\int d^d{\bf k} \, d\omega \, 
{\tilde \eta}({\bf k},\omega) \, e^{i({\bf k}\cdot {\bf r}-\omega t)} \; . 
\eeq 

In fact, by appling the 
Fourier transform ${\cal F}$ to Equation (\ref{linear-amari}) and using 
the properties of ${\cal F}$ with respect to derivatives and integrals 
(convolution theorem) one immediately finds 
\beq 
\left( - i \omega \tau \, + 1  -  
f'[u_0] {\tilde w}(k) \right) {\tilde \eta}({\bf k},\omega) = 0 \; ,  
\label{fourier-linear-amari}
\eeq
with $k=|{\bf k}|$. From this equation 
the dispersion relation $\omega=\omega_k$ 
between $\omega$ and $k$ reads 
\beq
\omega_k = {i\over \tau} \left( -1 + f'[u_0] {\tilde w}(k) \right) . 
\eeq

We stress that Equation (\ref{fourier-linear-amari}) can be rewritten as 
\beq 
G_0^{-1} \, {\tilde \eta}({\bf k},\omega) = 0 \; , 
\eeq
where 
\beq 
G_0 = {1\over - i \omega \tau \, + 1  -  
f'[u_0] {\tilde w}(k)  }
\eeq
is the Green function of the linearized Amari's equation.

Let us switch on the external stimulus, i.e.,  $s({\bf r},t)\neq 0$. 
The corresponding linearized Amari's equation in reciprocal 
wavevector-frequency domain becomes 
\beq 
G_0^{-1} \, {\tilde \eta}({\bf k},\omega) = {\tilde s}({\bf k},\omega) 
\eeq
from which we get the solution 
\beq 
{\tilde \eta}({\bf k},\omega) = G_0 {\tilde s}({\bf k},\omega) 
\eeq
namely 
\beq 
{\tilde \eta}({\bf k},\omega) = 
{{\tilde s}({\bf k},\omega) \over 
-i \omega \tau \, + 1  -  f'[u_0] {\tilde w}(k) } \; . 
\label{solution}
\eeq

The power spectrum $P_{\eta}({\bf k},\omega)$ 
of the neural perturbation $\eta({\bf k},t)$ is defined as 
\beq 
P_{\eta}({\bf k},\omega) = |{\tilde \eta}({\bf k},\omega)|^2 
\eeq
and taking into account Equation (\ref{solution}) it is given by 
\beq 
P_{\eta}({\bf k},\omega) = { |{\tilde s}({\bf k},\omega)|^2 \over 
\omega^2 \tau^2  + \left(1  -  f'[u_0] {\tilde w}(k) \right)^2 } 
\eeq

This simple but elegant analytical formula gives immediately the 
power spectrum of the neutral field knowing the Fourier 
transform ${\tilde s}({\bf k},\omega)$ of the external stimulus 
$s({\bf k},t)$. 

In the case of an instantaneous stimulus of amplitude $s_0$ localized 
at position ${\bf r}={\bf 0}$ and time $t=0$, i.e.,  
\beq 
s({\bf r},t) = s_0 \, \delta^{(d)}({\bf r}) \, \delta(t) \; , 
\eeq
with $\delta^{(d)}({\bf r})$ the Dirac delta function in $d$ dimensions, 
the power spectrum of the neural perturbation becomes 
\beq 
P_{\eta}(k,\omega) = { s_0^2 \over 
\omega^2 \tau^2  + \left(1  -  f'[u_0] {\tilde w}(k) \right)^2 } \; . 
\eeq 

\begin{figure}[b]
\centerline{\epsfig{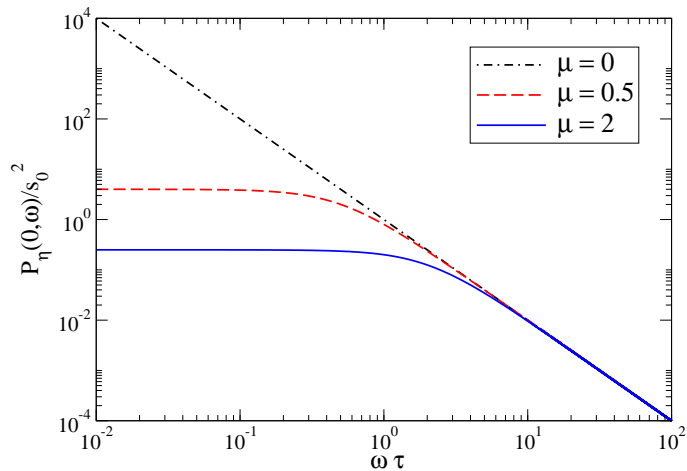}}
\small 
\caption{Scaled power spectrum $P_{\eta}(0,\omega)/s_0^2$ 
of the neutral field as a function of the scaled frequency $\omega \tau$ 
in the case of an instantaneous and localized stimulus. Three values 
of the parameter $\mu=1-f'[u_0]{\tilde w}(0)$: $\mu =0$ (dot-dashed line), 
$\mu=0.5$ (dashed line), $\mu=2$ (solid line).} 
\label{fig1}
\end{figure}

In Figure \ref{fig1} {we} plot the power spectrum at $k=0$, i.e.,  
\beq 
P_{\eta}(0,\omega) = { s_0^2 \over \omega^2 \tau^2  + \mu^2 } \; , 
\eeq
as a function of the frequency $\omega$ 
in the case of the instantaneous and localized stimulus for three 
values of the parameter 
\beq 
\mu=1-f'[u_0]{\tilde w}(0) \; . 
\eeq

The figure clearly shows that for large 
frequencies ($\omega \gg \mu/\tau$) the power 
spectrum is described by the power law  
\beq 
P_{\eta}(0,\omega) \simeq {s_0^2\over \tau^2} {1\over \omega^2} \; . 
\eeq 

As previously discussed, this result, that is valid in the regime of 
small wavenumbers ($k\simeq 0$), is consistent with the scaling 
laws found in measurements of electroencephalography (EEG) \cite{eeg1}. 
It is important to stress that, as written also in the introduction, 
EEG spectra are quite complex and display clear resonances 
(alpha, beta, etc) beyond the $1/\omega^2$ shoulder \cite{eeg2}. 
These nontrivial features can be captured by the inclusion of  
a time delay in the Amari equation~\cite{robinson,jirsa}. 

The $1/\omega^2$ power-law of the Amari equation, 
which is mapped into the Wilson-Cowan equation~\cite{review-bressloff}, is not surprising. In fact, 
the exponential decay ($\mu >0$) in the time domain, i.e.,  
\beq 
\eta(t) = \eta(0) \, e^{-\mu t/\tau} 
\eeq 
and consequently 
\beq 
u(t) = u_0 + \eta(0) \, e^{-\mu t/\tau} \; , 
\label{sbiro}
\eeq
implies a Lorentzian power spectrum in the frequency 
domain \cite{fourier-book}. Note that Equation (\ref{sbiro}) means that $u_0$, 
which satisfies Equation (\ref{constant}), is a stable fixed point 
of the uniform Amari equation 
\beq 
\tau {\partial \over \partial t} u(t) = - u(t) + W f[u(t)]
\label{pippe} 
\eeq
under the condition $\mu=1-f'[u_0]{\tilde w}(0)>0$.

\subsection{Diffusion Equation From the Linearized Amari Equation}

It is interesting to observe that for small wavenumbers $k$ 
we can write 
\beq 
{\tilde w}(k) \simeq {\tilde w}(0) + {1\over 2} {\tilde w}''(0) k^2 
\label{long}
\eeq
and the linearized Amari equation (\ref{fourier-linear-amari}) becomes 
\beq 
\left( - i \omega \tau \, + 1  -  
f'[u_0]  w(0) - {1\over 2} f'[u_0]  w''(0) k^2 \right) 
{\tilde \eta}({\bf k},\omega) = 0 \; .   
\eeq

Performing the Fourier antitransform \cite{fourier-book} of this equation 
we obtain 
\beq 
\tau \, {\partial \over \partial t} \eta({\bf r},t) = 
\left( D\, \nabla^2 - \mu \right) \eta({\bf r},t) \; , 
\label{diffusion}
\eeq
with $\mu=1-f'[u_0] {\tilde w}(0)$ and $D=-f'[u_0] {\tilde w}''(0)/2$. 
Equation (\ref{diffusion}) is a diffusion equation with $D>0$ the diffusion 
coefficient, which can also be formally interpreted as a time-dependent 
Schr\"odinger equation with imaginary time \cite{pde-book}, and it is 
clearly reliable only for large wavelengths $\lambda=2\pi/k$. 
It is well known the Equation (\ref{diffusion}) admits meaningful 
analytical solutions \cite{pde-book}. For instance, given the Gaussian initial 
condition 
\beq 
\eta({\bf r},0) = \eta_0 \, e^{-{r^2\over \sigma^2}} 
\label{in-gauss}
\eeq 
induced by some stimulus, the time-dependent 
solution of Equation (\ref{diffusion}) reads 
\beq 
\eta({\bf r},t) = { \eta_0 \over \zeta(t)^{d/2} } \, 
e^{-{r^2\over \sigma^2 \zeta(t)}} \, e^{-{\mu t/\tau}} 
\eeq
with 
\beq 
\zeta(t) = 1+{2D\over \sigma^2\tau}t \;  , 
\label{gdit}
\eeq
and consequently 
\beq 
u({\bf r},t) = u_0 + { \eta_0 \over \zeta(t)^{d/2} } \, 
e^{-{\mu t/\tau}} \, e^{-{r^2\over \sigma^2 \zeta(t)}} \; 
\eeq
is a solution of the Amari Equation (\ref{amari}) under the conditions 
$\eta_0\ll u_0$ (small perturbation) and $\sigma \gg \sqrt{D}$ 
(large wavelengths). 

\section{Reaction-Diffusion From the Amari Equation 
with Weak Nonlinearity}

Let us now consider the effect of a weak nonlinearity 
in the Amari equation. In particular, in the expansion of Equation (\ref{expand}) 
we add a quadratic term, namely 
\beq
f[u({\bf r}',t)] \simeq f[u_0] + f'[u_0] \eta({\bf r}',t) + 
{1\over 2} f''[u_0] \eta({\bf r}',t)^2 \; . 
\label{expand1}
\eeq

Taking into account also the small wavenumber (long wavelength) expansion 
of Equation (\ref{long}) we immediately find a nonlinear reaction-diffusion 
equation 
\beq 
\tau \, {\partial \over \partial t} \eta({\bf r},t) = 
\left( D\, \nabla^2 - \mu \right) \eta({\bf r},t) + 
g \, \eta({\bf r},t)^2 \; ,
\label{diffusion1}
\eeq
with $g=f''[u_0] {\tilde w}'(0)$ and neglecting 
the term proportional to $f''[u_0] {\tilde w}''(0)/4$  
that is at the second order in both $k$ and $\eta$.

It is well known that reaction-diffusion equations admit 
several kind of solutions: traveling waves, stripes, 
 and 
dissipative solitons \cite{pde-book}. Here we consider the 
case of uniform initial perturbation $\eta({\bf r},0)=\eta_0$ 
such that the time-dependent but spatially uniform solution 
$\eta(t)$ of Equation (\ref{diffusion1}) can be obtained from 
\beq 
\tau \, {\partial \over \partial t} \eta(t) = 
- \mu \, \eta(t) + g \, \eta(t)^2 \; . 
\label{diffusion2}
\eeq

By using separation of variables and integration 
from Equation (\ref{diffusion2}) we obtain the solution 
\beq 
\eta(t) = \eta_0 \, {e^{-\mu t/\tau} \over 1 + 
{\eta_0 g \over \mu } \left( e^{-\mu t/\tau}  - 1 \right) } \; , 
\label{piox}
\eeq

In the special case $\mu=0$ the uniform solution is 
\beq 
\eta (t) = {\eta_0 \over 1 - {\eta_0 g\over \tau} t } \; . 
\label{pioxi}
\eeq

\begin{figure}[b]
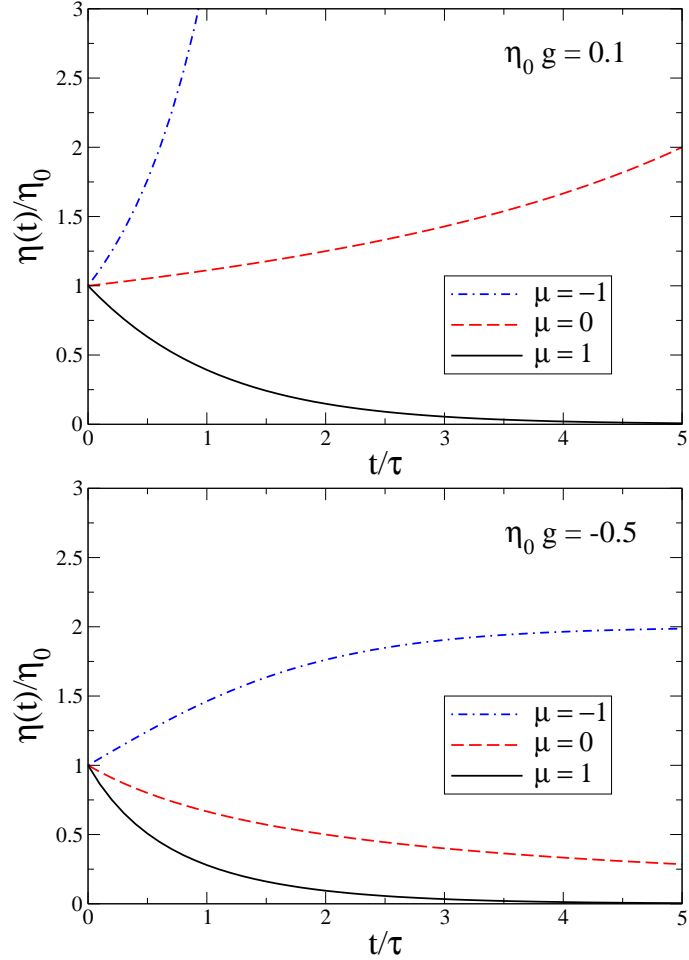

\centerline{\epsfig{file=power-f2.eps,width=9cm,clip=}}
\centerline{\epsfig{file=power-f3.eps,width=9cm,clip=}}
\small 
\caption{Time evolution of the scaled perturbation $\eta(t)/\eta_0$ 
of the neutral field as a function of the scaled time $t/\tau$ 
in the case of a uniform initial perturbation $\eta_0$ with respect 
to a uniform background $u_0$. Curves obtained from Equations (\ref{piox}) 
and (\ref{pioxi}) with three values of $\mu$: dot-dashed line with 
$\mu=-1$, dashed line with $\mu=0$, solid line with $\mu=1$.  
Upper panel: $\eta_0\, g=0.1$. Lower panel $\eta_0\, g=-0.5$.} 
\label{fig2}
\end{figure}

The time evolution of $\eta(t)$ is shown for three values of $\mu$ 
in Figure \ref{fig2}. The upper panel of Figure \ref{fig2} clearly shows 
that, chosing $\eta_0\, g =0.1$, for $\mu=1$ there 
is an exponential decay to zero, for $\mu=0$ there is a polynomial growth, 
and for $\mu=-1$ an exponential growth. Actually, for $\eta_0\, g>0$ 
it follows that $\eta(t)$ diverges at $t=\tau/(\eta_0 g)$ for $\mu=0$ 
and at $t=(\tau/|\mu|)\ln{(1+(|\mu|/(\eta_0g)))}$ for $\mu <0$. 
Obviously, Equation (\ref{diffusion}) is valid 
if the perturbation 
$\eta({\bf r},t)$ is small, thus a solution makes sense only perturbatively  
and Equations (\ref{piox}) and (\ref{pioxi}) can be trusted 
only for $|\eta(t)| \ll |u_0|$. However, if $\eta_0\, g <0$ 
there are no divergences: For $\mu>0$ there is an exponential decay to zero, 
for $\mu=0$ there is polynomial decay to zero, and for $\mu<0$ one finds 
$\eta(t) \to \mu/g$ as $t\to \infty$. 
These trends are explicitly shown in the lower panel of Figure \ref{fig2}.

\subsection{Dissipation and Neural Action}

In this subsection we analyze the dissipative nature of 
Equation (\ref{diffusion1}). Remarkably, dissipative 
equations can be derived from a variational principle 
by doubling the degees of freedom \cite{bateman,vitiello}. 
Let us show this interesting property by introducing 
the dual field $\phi({\bf r},t)$ and the neural action functional 
\beq 
S[\eta({\bf r},t),\phi({\bf r},t)] = \int d^d{\bf r}\, dt \ {\cal L} \; , 
\label{action}
\eeq
where 
\beq 
{\cal L} = \phi({\bf r},t) \left( \tau {\partial\over\partial t} 
- D \nabla^2 + \mu \right) \eta({\bf r},t) - 
g \, \phi({\bf r},t) \eta({\bf r},t)^2 
\label{lagrangian}
\eeq
is the Lagrangian density of the neural field. It is 
straightforward to see that the Euler-Lagrange equation 
\beq
{\partial {\cal L}\over\partial \phi} = 0 
\label{abo}
\eeq
obtained extemizing the neural action (\ref{action}) 
with respect to the 
dual field $\phi({\bf r},t)$ gives exactly Equation~(\ref{diffusion}), 
while the Euler-Lagrange equation 
\beq
{\partial {\cal L}\over\partial \eta} - 
{\partial \over\partial t} {\partial {\cal L}\over\partial 
(\partial_t \eta)} - \nabla \cdot {\partial {\cal L}\over\partial 
(\nabla \eta)} + \nabla^2 {\partial {\cal L}\over\partial 
(\nabla^2 \eta)} = 0 
\label{bbo}
\eeq
obtained extremizing the neural action (\ref{action}) 
with respect to the neutral field $\eta({\bf r},t)$ gives the 
differential equation of the dual field $\phi({\bf r},t)$, that is 
\beq
-\tau \, {\partial \over \partial t} \phi({\bf r},t) =
\left( D\, \nabla^2 + \mu \right) \phi({\bf r},t) + 
2 \, g \, \eta({\bf r},t) \, \phi({\bf r},t) \; . 
\label{dual-diffusion1}
\eeq 

Notice that Equation (\ref{diffusion}) is independend of $\phi({\bf r},t)$ while 
Equation (\ref{dual-diffusion1}) depends on $\eta({\bf r},t)$. 

The neural action functional 
(\ref{action}) with (\ref{lagrangian1}) is similar to the 
renormalized neural action introduced in Ref. \cite{buice2}) 
within a stochastic extension of the Wilson-Cowan model. 
Using our notations the Lagrangian density of Ref. \cite{buice2} reads 
\beq 
{\cal L} = \phi({\bf r},t) \left( \tau {\partial\over\partial t} 
- D \nabla^2 + \mu \right) \eta({\bf r},t) 
- g \, \left( \phi({\bf r},t) \eta({\bf r},t)^2 - \phi({\bf r},t)^2  
\eta({\bf r},t) \right) \; ,  
\label{lagrangian1}
\eeq
which clearly gives a quite different reaction-diffusion equation 
where it appears explicitly the dual field $\phi({\bf r},t)$ 
that induces a stochastic noise in the dynamics 
of the neural field $\eta({\bf r},t)$. 
As discussed in Ref. \cite{buice1,buice2} 
the Lagrangian density (\ref{lagrangian1}) 
is called Reggeon field theory \cite{gribov} 
and it has a directed percolation phase transition 
when $\mu =0$ \cite{cardy}, i.e.,  when one cross the 
stability region of the uniform configuration 
(that is dynamically stable only for $\mu>0$, see Equation (\ref{pippe})). 

In our approach the dual field 
$\phi({\bf r},t)$ does not influence the deterministic 
dissipative dynamics of the neural field, given by 
Equation (\ref{diffusion1}). However, $\phi({\bf r},t)$ is necessary 
to derive Equation (\ref{diffusion1}) from a variational principle.

\section{Conclusions}

By analyzing the Amari equation of a neural field 
we have obtained an analytical formula for its power spectrum 
under the assumption of a small perturbation around a stationary 
uniform neural field in the presence of a generic external stimulus. 
In the case of a istantaneous and localized external stimulus 
the power spectrum is quite simple and for large frequencies it 
scales as $1/\omega^2$. It is important to observe that also 
in \cite{plus1} there is an explicit derivation 
of $1/\omega^2$ power law for voltage in cable equations, 
which are diffusion equations 
with a source term, while in \cite{plus2} the same power law is derived 
analytically for large frequencies in a class of integro-differential 
equation models with various types of local connectivities. 
In this paper we have also shown that for large wavelengths (small wavenumbers) 
the linearized Amari equation is equivalent to a diffusion equation, 
for which we write the space-time dependent analytical solution in the case 
of a Gaussian initial perturbation. 
Finally, taking into account quadratic corrections  
to the linearized Amari equation we have deduced 
a reaction-diffusion equation, which can be formally derived by
extremizing a neural action functional. 
This neural action is similar to the one proposed in \cite{buice2} on the basis of a stochastic extension  
of the Wilson-Cowan model. We have also shown that, 
for some specific initial conditions, our reaction-diffusion 
equation admits meaningful spatially uniform analytical solutions.   \vspace{6pt}

The author thanks F. Sattin and F. Toigo for useful discussions.

\end{document}